\documentclass[sigconf]{acmart}

\usepackage{packages}
\AtBeginDocument{%
  }

\setcopyright{acmlicensed}
\copyrightyear{2026}
\acmYear{2026}
\acmDOI{XXXXXXX.XXXXXXX}
\acmConference[Pre-print]{}{}{}

\acmISBN{978-1-4503-XXXX-X/2026/06}

\begin{document}

\title{\green{Green Prompt Engineering}: Investigating the Energy Impact of \\Prompt Design in Software Engineering}


\author{Vincenzo De Martino$^{*}$ $^{\ddagger}$, Mohammad Amin Zadenoori$^{\dagger}$, Xavier Franch$^{\ddagger}$, Alessio Ferrari$^{\S}$}

\affiliation{%
  $^{*}$Software Engineering (SeSa) Lab, University of Salerno \country{Italy}, vdemartino@unisa.it
}

\affiliation{%
  $^{\dagger}$University of Padova \country{Italy}, Amin.zadenoori@unipd.it}

\affiliation{%
  $^{\ddagger}$Universitat Politècnica de Catalunya \country{Spain}, \{vincenzo.de.martino, xavier.franch\}@upc.edu
}

\affiliation{%
  $^{\S}$University College Dublin (UCD) \country{Ireland}, alessio.ferrari@ucd.ie
}


\renewcommand{\shortauthors}{De Martino et al.}

\begin{abstract}
Language Models are increasingly applied in software engineering, yet their inference raises growing environmental concerns. Prior work has examined hardware choices and prompt length, but little attention has been paid to linguistic complexity as a sustainability factor. This paper introduces \green{Green Prompt Engineering}, framing linguistic complexity as a design dimension that can influence energy consumption and performance. We conduct an empirical study on requirement classification using open-source Small Language Models, varying the readability of prompts. Our results reveal that readability affects environmental sustainability and performance, exposing trade-offs between them. For practitioners, simpler prompts can reduce energy costs without a significant F1-score loss; for researchers, it opens a path toward guidelines and studies on sustainable prompt design within the \green{Green AI} agenda.
\end{abstract}




\maketitle

\section{Introduction}
\label{sec:intro}
The exponential use of language models (LMs) has driven major advances in natural language processing (NLP) and software engineering (SE) \cite{hou2024large,zheng2025towards,fan2023large}, but has also raised serious concerns on the environmental sustainability of the training and inference phases \cite{bender2021dangers,strubell2020energy,shi2024greening}. Although the emissions from a single inference may seem negligible, their cumulative impact can surpass that of training when models are queried at scale \cite{stanford2024Index}. Even minor additions to prompts, such as \quoted{hello} or \quoted{thanks}, can significantly increase energy when multiplied across billions of queries \cite{nytimesSayingThank}, and it is estimated that a single ChatGPT query may require up to half a liter of water for cooling \cite{EveryTime}.

Prompt engineering emerged as a discipline for developing and optimizing prompts, enabling the efficient use of LMs across various applications and research topics \cite{vogelsang2025impact,della2025prompt,de2025developers}. The prompt, as the interface between the user and the model, plays a fundamental role, as it directly influences the model's behavior, the quality of responses, and the resources to produce an output, and thus its energy consumption.  
Community initiatives have begun to offer interactive tools such as ChatUI Energy\footnote{\url{https://huggingface.co/spaces/jdelavande/chat-ui-energy}}, to provide real-time energy estimates and visualize the environmental costs of LM interactions.

Prior research on prompting has examined environmental sustainability from multiple perspectives, including the relationship between energy consumption and prompt length or formulation \cite{husom2024price}, design choices \cite{rubei2025prompt}, and the adoption of Small Language Models (SLMs) for energy-efficient code generation \cite{ashraf2025greencodepromptingsmall}. Other works explored methods to limit token generation \cite{guo2024stop}, and non-essential formatting tokens raising costs without accuracy gains \cite{pan2025hidden}.

Analogous to the concept of \green{Green AI} \cite{schwartz2020green}, in this paper, we introduce \green{Green Prompt Engineering}, which we define as the portfolio of prompt design practices that aim to maintain or improve performance while reducing the environmental sustainability of inference.
In contrast, \red{Red Prompt Engineering}, like \red{Red AI} \cite{schwartz2020green}, prioritizes performance alone, overlooking the associated energy, carbon, and water footprint. 

Evaluating prompts not only on performance but also on energy consumption encourages sustainable prompt design practices, with positive environmental impacts.
In line with these concepts, we aim to initiate a community discussion and raise awareness of the sustainability implications of prompt design, facilitating collective progress toward energy-efficient LM usage. For this reason:
\idea{\faLightbulb[regular] \hspace{0.05cm}
Our idea is to explore how variations in prompt linguistic complexity affect the energy consumed by an LM during inference and its trade-offs with performance, establishing a basis for \green{Green Prompt Engineering} as a discipline.}

To ground this idea, we focus on requirement classification using open-source LMs, a well-established task in SE~\cite{ferrari2025formal,alhoshan2025effective,almonte2025automated,10.1007/978-3-031-88531-0_15}. This task provides a controlled setting to assess the impact of prompt complexity and ensure relevance to the field. This work presents the first empirical study of linguistic complexity as a sustainability dimension in prompt engineering.
\replicationpackagebox{To ensure verifiability and replicability, we provide a replication package with all raw data, the dataset, prompts, and analysis scripts~\cite{appendix}.}

\section{Study Design}
\label{sec:design}
To define our research goal, we follow the Goal-Question-Metric (GQM) approach \cite{caldiera1994goal}.
The \emph{goal} of this study is to conduct a \textit{preliminary} investigation of how the linguistic complexity of prompts influences the energy consumption of language models. The \emph{purpose} is to provide empirical evidence that can guide the adoption of \green{Green Prompt Engineering} practices. The \emph{context} is requirements classification performed with open-source SLMs. The \emph{perspective} is twofold:
(i) practitioners can leverage the findings to design prompts that balance performance and energy efficiency;
(ii) researchers can build on this evidence to explore sustainable prompt optimization strategies. From this goal, we derive two research questions:
\sterqbox{RQ\textsubscript{1} – Energy Consumption}{How does prompt linguistic complexity influence the energy consumption of SLM inference?}
\sterqbox{RQ\textsubscript{2} – Trade-Offs}{What trade-offs exist between the energy consumption of prompts and SLM performance measured with F1-score?}


In designing and reporting our study, we employed the guidelines (i) by Wohlin et al.~\cite{wohlin2012experimentation}, (ii) by the \textsl{ACM/SIGSOFT Empirical Standards}\footnote{Available at: \url{https://github.com/acmsigsoft/EmpiricalStandards}. Given the nature of our study and the currently available standards, we followed the \quoted{\textsl{General Standard}}, and \quoted{\textsl{Benchmarking}} guidelines.}, (iii) the Evaluation Guidelines for Empirical Studies in Software Engineering involving LLMs \cite{baltes2025evaluation}, and (iv) the PRIMES 2.0 prompt design framework \cite{de2025methodological}.

\subsection{Objects of the Study}
\label{sec:objects}
\textbf{Dataset.} The task under investigation is the binary classification of software requirements, a widely studied area in requirements engineering (RE) ~\cite{zhao2021natural,zadenoori2025largelanguagemodelsllms,almonte2025automated,kurtanovic2017automatically}. Distinguishing requirements categories is particularly relevant as this facilitates maintenance, reuse, and team allocation~\cite{alhoshan2023zero,bashir2023requirements}. 
We used the PROMISE dataset \cite{boetticher2007promise}, a benchmark of 625 requirements from 15 documents across multiple domains, manually labeled as Functional or Non-functional and it provides a balanced, reliable classification baseline \cite{hu2025assessing,almonte2025automated,cleland2007automated}.

\begin{table}[h]
\caption{Flesch categories with prompts and characteristics.}
\label{table:flesch_reading}
\centering
\footnotesize
\rowcolors{1}{gray!15}{white}
\begin{tabular}{|c|c|c|c|c|}
\rowcolor{black}
\textcolor{white}{\textbf{Score Range}} &
\textcolor{white}{\textbf{Prompt}} &
\textcolor{white}{\textbf{Flesch}} &
\textcolor{white}{\textbf{\#Words}} &
\textcolor{white}{\textbf{\#Tokens}} \\
\hline

100.0-90.0 - \textbf{5th grade}  & P1 & 93.11 & 75 & 99 \\
\cline{2-5}
 (Very easy to read) & P2 & 92.99 & 70 & 94 \\
\hline

90.0-80.0 - \textbf{6th grade} & P3 & 84.14 & 78 & 106 \\
\cline{2-5}
 (Easy to read) & P4 & 83.53 & 80 & 103 \\
\hline

80.0-70.0 - \textbf{7th grade} & P5 & 74.64 & 67 & 93 \\
\cline{2-5}
(Fairly easy to read) & P6 & 70.25 & 128 & 167 \\
\hline

70.0-60.0 - \textbf{8th \& 9th grade} & P7 & 61.45 & 199 & 256 \\
\cline{2-5}
(Plain English) & P8 & 60.14 & 209 & 273 \\
\hline

60.0-50.0 - \textbf{10th to 12th grade} & P9 & 55.94 & 184 & 243 \\
\cline{2-5}
(Fairly difficult to read) & P10 & 52.83 & 191 & 250 \\
\hline

50.0-30.0 - \textbf{College} & P11 & 46.16 & 161 & 220 \\
\cline{2-5}
(Difficult to read) & P12 & 34.94 & 154 & 181 \\
\hline

30.0-10.0 - \textbf{College graduate}  & P13 & 28.62 & 121 & 178 \\
\cline{2-5}
(Very difficult to read) & P14 & 18.15 & 146 & 192 \\
\hline

10.0-0.0 - \textbf{Professional} & P15 & 2.51 & 146 & 206 \\
\cline{2-5}
(Extremely difficult to read) & P16 & 0.94 & 96 & 131 \\
\hline

\end{tabular}
\end{table}

\begin{figure}[h]
    \centering
    \includegraphics[width=0.6\linewidth]{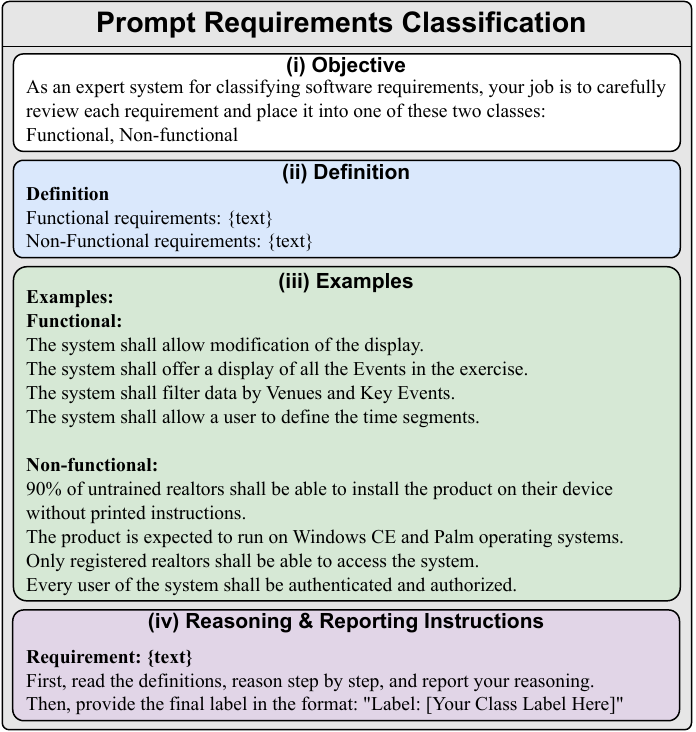}
    \caption{Prompt example to classify Requirements.}
    \label{fig:prompt}
\end{figure}

\textbf{Prompt Configurations.} We adopted a \emph{Chain-of-Thought (CoT) + few-shot} prompting, where the examples included the input and label but also an explicit reasoning step \cite{kojima2022large,
wei2022chain,kojima2022large}. Previous studies identified this prompting technique as the best performing one for  requirements classification~\cite{10.1007/978-3-031-88531-0_15}. This design encourages the model to mimic step-by-step reasoning before producing the final classification, shown in Figure \ref{fig:prompt}, structured into four components to achieve the study objective.
Across all prompt variants, only component (ii)—the requirements definitions—was modified to adjust linguistic complexity based on the \emph{Flesch Reading Ease} score \cite{kincaid1975derivation}, computed with \texttt{textstat} from average sentence length and syllables per word, capturing syntactic and lexical difficulty

This score ranges from 0 (very difficult) to 100 (very easy), and we partitioned into eight readability groups in Table~\ref{table:flesch_reading}; for each range, we generated two prompts, yielding 16 variants that differed only in definition complexity while preserving the original semantic meaning.  Prompts were produced from original definitions derived from the literature~\cite{4384163} using ChatGPT 5 with instructions to vary sentence length and lexical choices while keeping semantics intact; the first author generated them and the others reviewed them \cite{de2025methodological}. 

\subsection{Subjects of the Study}
\label{sec:subjects}



\textbf{Models.} We selected open-source SLMs rather than proprietary LLMs because: \emph{(i)} open-source LMs allow local deployment with full control over execution and monitoring, ensuring transparent energy measurements and reproducibility \cite{de2025methodological,baltes2025evaluation}; \emph{(ii)} SLMs reduce memory and energy demands while balance between efficiency, accessibility, and performance, making them a more sustainable choice than LLMs \cite{wang2024comprehensive}. Table~\ref{table:llm_specs} describe  model characteristics.

\begin{table}[h]
\caption{Open-source SLMs (temperature = 0).}
\label{table:llm_specs}
\centering
\footnotesize
\rowcolors{1}{gray!15}{white}
\begin{tabular}{|c|c|c|c|p{0.8cm}|}
\rowcolor{black}
\textcolor{white}{\textbf{Organization}} & 
\textcolor{white}{\textbf{Model}} & 
\textcolor{white}{\textbf{Params}} &  
\textcolor{white}{\textbf{Release}} & 
\textcolor{white}{\textbf{Context Length}} \\
\hline
Qwen & \href{https://huggingface.co/Qwen/Qwen2-7B-Instruct}{Qwen2-7B-Instruct} & 7.62B & Jul 2024 & 32k \\
\hline
TII & \href{https://huggingface.co/tiiuae/Falcon3-7B-Instruct}{Falcon3-7B-Instruct} & 7.46B & Dec 2024 & 32k \\
\hline
IBM & \href{https://huggingface.co/ibm-granite/granite-3.2-8b-instruct}{granite-3.2-8b-instruct} & 8.17B & Feb 2025 & 131k \\
\hline
Mistral & \href{https://huggingface.co/mistralai/Ministral-8B-Instruct-2410}{Ministral-8B-Instruct-2410} & 8.02B & Oct 2024 & 128k \\
\hline
Meta LLaMA & \href{https://huggingface.co/meta-llama/Meta-Llama-3-8B-Instruct}{Meta-Llama-3-8B-Instruct} & 8.03B & Apr 2024 & 8k \\
\hline
\end{tabular}
\end{table}
\textbf{Metrics.} We evaluated each SLM along two dimensions. For \textbf{RQ$_1$}, we measured \textbf{inference energy consumption} using \texttt{CodeCarbon}\footnote{\url{https://codecarbon.io/}}, which standardizes CPU/RAM (RAPL) and GPU (\texttt{pynvml}) monitoring, reporting all values in kilojoules (kJ) for cross-run comparability \cite{rubei2025prompt,de2025examining}. For \textbf{RQ$_2$}, we measured \textbf{F1-score} macro averaged, a standard metric in requirements classification and NLP \cite{almonte2025automated,rejithkumar2025nice}. Energy captures the environmental cost of prompts, while F1-score reflects output quality, enabling analysis of trade-offs. 

\subsection{Experiment, Data Collection and Analysis}
\label{sec:experiment}
After defining the experimental subjects and objects, we designed our experiment and outlined how to collect and analyze data.

\textbf{Experimental setup and Execution.}
All experiments were conducted on August 27, 2025, on a Linux 6.14 server with an Intel i9-13900K CPU, 128 GB RAM, an NVIDIA RTX 4090 GPU, and Python 3.12. We evaluated five open-source SLMs with 16 prompts each (eight range levels × two variants), totaling 80 configurations. To reduce non-determinism from hardware/software variability, each configuration was repeated 10 times, yielding 800 runs that required seven days, two hours, and 14 minutes.

Following Cruz’s guidelines \cite{cruz2021green}, we terminated unnecessary background processes, ran a five-minute Fibonacci sequence to warm up the CPU, and inserted a one-minute pause between runs. Metrics were automatically logged to \texttt{.csv} files using CodeCarbon.


\textbf{Data collection and Analysis.} 
For each execution over the entire dataset, we collected (i) energy consumption for \textbf{RQ\textsubscript{1}}, and (ii) SLM predictions, which were then used to compute the F1-score with (i) for \textbf{RQ\textsubscript{2}}. For \textbf{RQ\textsubscript{1}}, we first computed the average energy consumption across score ranges. To separate the effect of prompt length from ranges, we used a linear mixed-effects model \cite{baltagi2008econometric} with SLM as random effect and \# words and groups as fixed effects. Normality was then assessed with Q–Q plots and the Shapiro–Wilk test ($\alpha=0.05$) \cite{shapiro1965analysis}. As assumptions were not always met, we applied the Friedman test \cite{sprent2007applied}, followed by the Nemenyi post-hoc test \cite{nemenyi1963distribution,hollander2013nonparametric}. To control family-wise error, we applied the Holm–Bonferroni correction \cite{holm1979simple}. In addition, we applied Cliff's delta ($\delta$), a nonparametric effect size for pairwise comparisons \cite{cliff1993dominance}.

For \textbf{RQ\textsubscript{2}}, we quantified the association between energy consumption and F1 using the nonparametric correlation tests Spearman's $\rho$ \cite{spearman1961proof} and Kendall's $\tau$ \cite{kendall1938new} for each model.

\section{Preliminary Results}
\subsection{RQ$_1$ - Energy Consumption}
\begin{figure}[h]
    \includegraphics[width=0.9\linewidth]{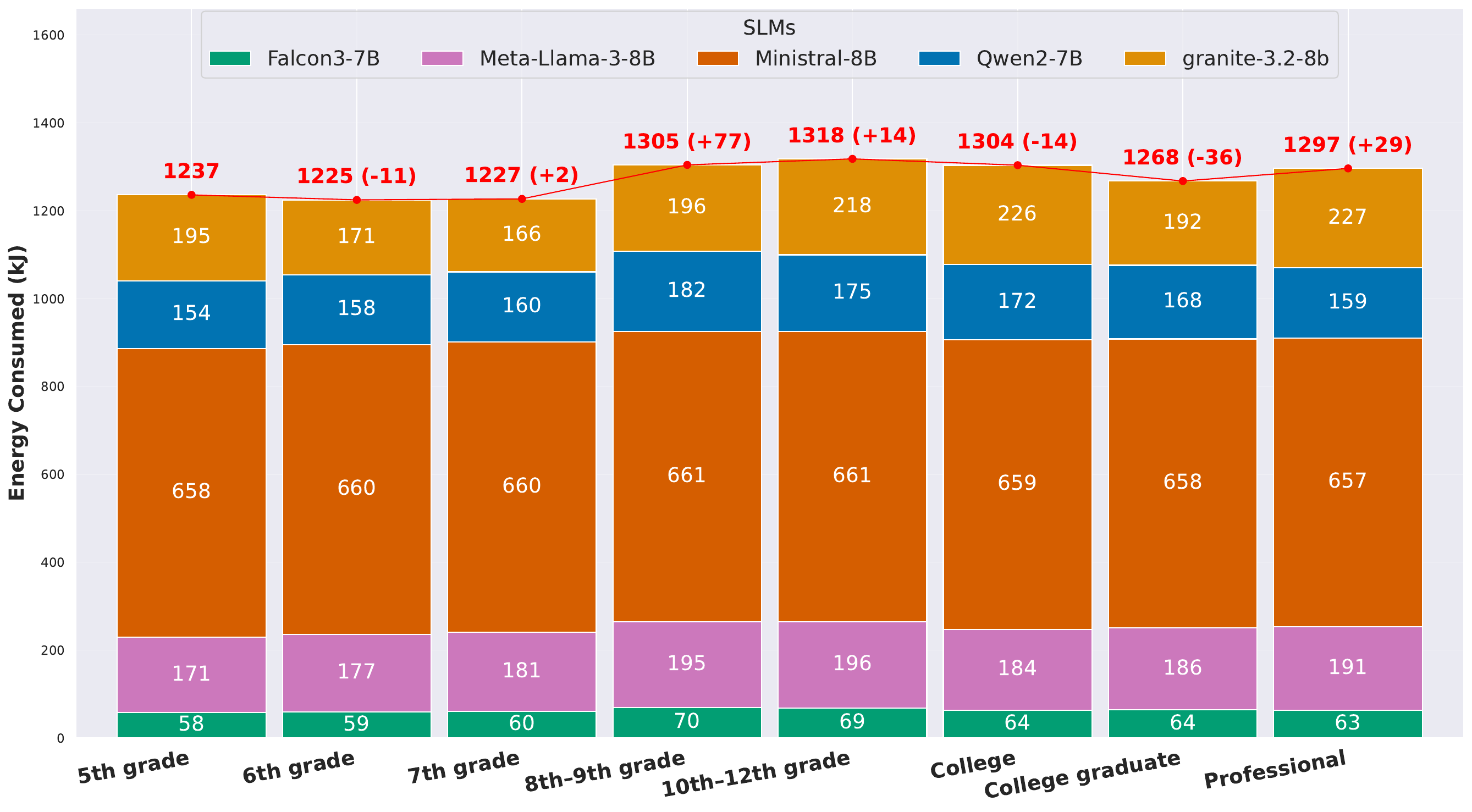}
    \caption{Average energy consumption across score ranges.}
    \label{fig:energy_categories_models}
\end{figure}

Figure~\ref{fig:energy_categories_models} shows how prompt complexity affects energy consumption across SLMs. Prompts written at simpler linguistic levels (e.g., \emph{5th grade}) generally require less energy, while more complex ones (especially \emph{10th-12th grade}) require more energy. 
Although this pattern is observable across all models, there are differences between models even if they have a similar \# parameters. For example, models such as \emph{granite-3.2-8b} and \emph{Meta-Llama-3-8B} show higher energy consumption than models \emph{Qwen2-7B} and \emph{Falcon-3-7B}. 

\begin{figure}[h]
    \includegraphics[width=0.7\linewidth]{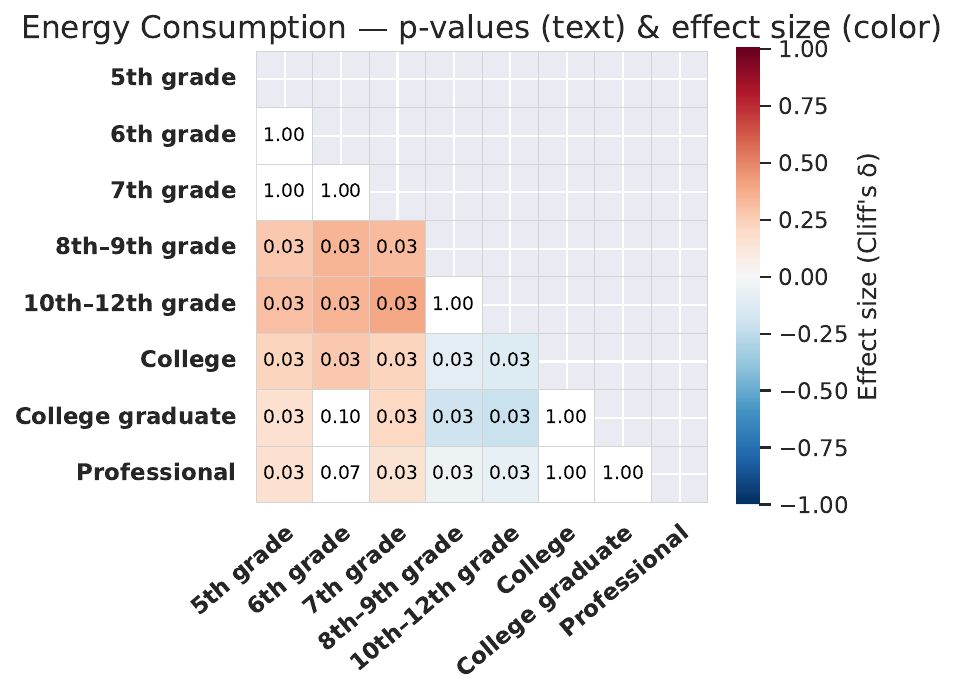}
    \caption{Statistical Differences in Energy Consumption.}
    \label{fig:energy_stat_differences}
\end{figure}

To further examine these differences, Figure~\ref{fig:energy_stat_differences} shows the results of statistical comparisons between pairs. The figure, with all the observations aggregated across the five SLMs, shows that many comparisons are statistically significant (\textit{p-value} < 0.05), particularly between low-complexity prompts (e.g., \emph{5th–7th grade}) and high-complexity prompts with higher effect size. In contrast, adjacent categories often do not differ significantly and have a lower effect size, suggesting a gradual rather than sudden increase in energy. 

Figure \ref{fig:energy_categories_models} shows that, across all models, the highest consumption occurred in the prompts \emph{8th-9th grade} and \emph{10-12th grade}. One possible explanation is that these prompts contain more words and tokens (Table~\ref{table:flesch_reading}), requiring greater processing time. To validate this hypothesis, Table~\ref{table:mixed_model} shows the mixed model results, where prompt length has a strong positive effect, while range groups display less differences. 
Overall, the prompt length is the dominant factor.
\begin{table}[h]
\caption{Mixed Linear Model (baseline = 5th grade).}
\label{table:mixed_model}
\centering
\footnotesize
\rowcolors{1}{gray!15}{white}
\begin{tabular}{|c|c|c|c|}
\rowcolor{black}
\textcolor{white}{\textbf{Predictor}} &
\textcolor{white}{\textbf{Coef.}} &
\textcolor{white}{\textbf{z}} &
\textcolor{white}{\textbf{p-value}} \\
\hline
Intercept & 237.511 & 2.30 & 0.022 \\
\hline
Word\_Count & 0.135 & 4.98 & $<$0.001 \\
\hline
6th grade & -3.173 & -1.96 & 0.050 \\
\hline
7th grade & -5.211 & -2.98 & 0.003 \\
\hline
8th--9th grade & -4.126 & -1.05 & 0.292 \\
\hline
10th--12th grade & 0.854 & 0.24 & 0.808 \\
\hline
College & 2.017 & 0.72 & 0.473 \\
\hline
College graduate & -1.943 & -0.84 & 0.400 \\
\hline
Professional & 5.485 & 2.64 & 0.008 \\
\hline
\end{tabular}
\end{table}

\rqanswer{1}{Our findings shows that model is the dominant factor, with \emph{Falcon-3-7B} the most efficient and \emph{Ministral-8B} the least. Linguistic complexity also influences SLM energy consumption, with higher prompt complexity and especially longer prompts consuming more than lower.}


\subsection{RQ$_2$ - Trade-Offs}

\begin{figure}[h]
    \includegraphics[width=0.95\linewidth]{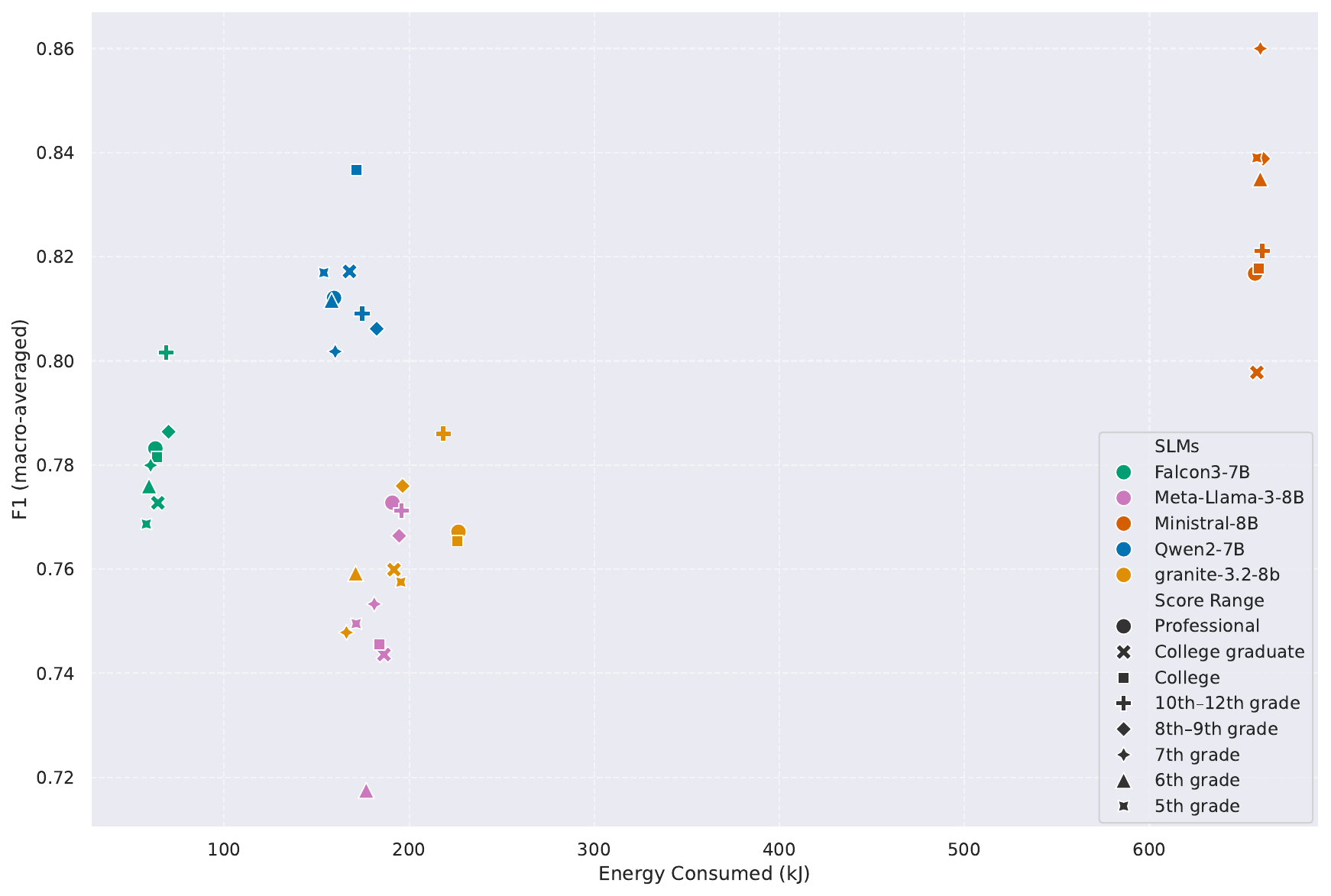}
    \caption{Scatterplot Average Energy Consumption and F1.}
    \label{fig:rq2_tradeoffs}
\end{figure}

Figure~\ref{fig:rq2_tradeoffs} shows the trade-offs between energy consumption and F1-score across ranges and SLMs. Higher linguistic complexity generally requires more energy, while F1 gains are modest and often stabilize at intermediate levels. \emph{Falcon-3-7B} remains the most efficient, keeping stable F1 with low energy. \emph{Meta-Llama-3-8B} shows a moderate rise in both metrics, while \emph{Ministral-8B} shows the highest F1 and energy. \emph{Qwen2-7B} improves at higher complexities with an energy peak at mid levels, whereas \emph{Granite-3.2-8B} consumes more at higher complexities with small F1 change. Notably, the most complex prompts never improved F1 nor efficiency, showing that excessive complexity adds cost without benefit.

To quantify the trade-off, we tested the monotonic association between energy consumption (Table \ref{table:correlation_energy_f1}). Results show moderate positive correlations for \emph{Falcon-3-7B}, \emph{Meta-Llama-3-8B}, \emph{Ministral-8B}, and \emph{Granite-3.2-8B}, while \emph{Qwen2-7B} showed no significant correlation. Overall, the trade-off of these metrics is model-dependent.

\begin{table}[h]
\caption{Correlation Energy Consumption and F1-score.}
\label{table:correlation_energy_f1}
\centering
\footnotesize
\rowcolors{1}{gray!15}{white}
\begin{tabular}{|c|c|c|c|c|}
\rowcolor{black}
\textcolor{white}{\textbf{Model}} &
\textcolor{white}{\textbf{Spearman $\rho$}} &
\textcolor{white}{\textbf{p-value}} &
\textcolor{white}{\textbf{Kendall $\tau$}} &
\textcolor{white}{\textbf{p-value}} \\
\hline
All Models & 0.2412 & $<$0.001 & 0.1833 & $<$0.001 \\
\hline
Falcon-3-7B & 0.5403 & $<$0.001 & 0.3710 & $<$0.001 \\
\hline
Meta-Llama-3-8B & 0.4212 & $<$0.001 & 0.3299 & $<$0.001 \\
\hline
Ministral-8B & 0.4472 & $<$0.001 & 0.3147 & $<$0.001 \\
\hline
Qwen2-7B & 0.0358 & 0.654 & -0.0178 & 0.745 \\
\hline
Granite-3.2-8B & 0.4403 & $<$0.001 & 0.2914 & $<$0.001 \\
\hline

\end{tabular}
\end{table}

\rqanswer{2}{Our results show that prompt complexity increases energy while F1 improves only up to medium levels, making excessive complexity a cost without benefit.}





\section{Threats to Validity}
\label{sec:threats}
\textbf{Construct validity.} Energy consumption was measured with CodeCarbon, which integrates RAPL and NVIDIA \texttt{pynvml}. Although widely used in prior studies \cite{rubei2025prompt,de2025examining}, it may underestimate consumption compared to hardware-level monitoring. Prompt complexity was measured with the Flesch index \cite{kincaid1975derivation}, capturing readability but not all features, and grouped into categories rather than exact scores, which may obscure finer differences. 


\textbf{Internal validity.} Background processes and system variability can influence energy measurements. To mitigate this, we followed established protocols \cite{cruz2021green,de2025methodological,baltes2025evaluation}, including warm-up runs, process control, and fixed configurations, though some noise is unavoidable.  

\textbf{External validity.} Our findings rely on requirements' classification \cite{boetticher2007promise}, one prompting strategy, and a limited selection of open-source SLMs and dataset. They may not generalize to other tasks, datasets, prompts, or larger models.

\textbf{Conclusion validity.} To ensure robustness, we applied appropriate statistical tests for distribution analysis and group comparisons, including corrections for multiple comparisons, reducing the likelihood of spurious results. Each prompt was run 10 times to account for variability; more runs could further increase stability.

\section{Discussion, Outlook, and Future Work}
\label{sec:discussion}

\begin{description}[leftmargin=0.3cm]

    \smallskip
    \item[\textbf{Impact of Prompt Linguistic Complexity}.] 
    We observe a hierarchy of effects with model architecture dominates, followed by prompt length, while lexical and syntactic variation have a subtler yet relevant impact. \#Word explain most differences, but complexity can still influence efficiency in specific ranges. \faBriefcase~\textbf{Practice.} Prompt engineering must evaluate performance and sustainability, but the choice of model remains the main factor. \faBook~\textbf{Research.} Future work should separate length from complexity to clarify their relative influence and consider alternative indices beyond Flesch \cite{kincaid1975derivation}.


    \item[\textbf{Cumulative Energy Effects}.] While training is typically episodic, prompting runs continuously at scale, making its cumulative energy cost substantial if not controlled. Even short queries can accumulate into significant emissions \cite{stanford2024Index}. Tools such as MLflow GenAI\footnote{\url{https://mlflow.org/docs/latest/genai/}} with CodeCarbon make these cumulative effects visible in real-world scenarios by tracking and showing emissions at the prompt and model level \cite{rubei2025prompt}. \faBriefcase~\textbf{Practice.} Practitioners must consider that frequent prompting can escalate costs rapidly and should adopt prompts and models proportionate to their needs. \faBook~\textbf{Research.} Further studies should capture the cumulative footprint of inference, and empirical work should systematically report CO\textsubscript{2} emissions over the long term.

    \smallskip
    \item[\textbf{Towards Guidelines for \green{Green Prompt Engineering}}.] 
    Prompt design is a sustainability lever accessible to all users, and its cumulative impact is substantial. The SE community should move beyond isolated studies toward shared principles and guidelines. 

    \faBriefcase~\textbf{Practice.} Checklists and tools can support practitioners in evaluating prompts for clarity, accuracy, and efficiency. Embedding sustainability at the human–model interface would raise awareness that each interaction contributes to the overall footprint of LM-based systems.
    \faBook~\textbf{Research.} Future work should conduct empirical guidelines across datasets, tasks, and LMs to consolidate \green{Green Prompt Engineering} as part of \green{Green AI} \cite{schwartz2020green}.

\end{description}

\section*{Acknowledgments.} The carbon footprint of our experiment is approximately 18.7 Kg CO\textsubscript{2}, comparable to powering an average electric car for 546.48 km (339.57 miles) \cite{zhao2023quantifying}. Chat-GPT5 was used to improve the readability, and the authors remain fully responsible for the final content. 


\bibliographystyle{ACM-Reference-Format}
\bibliography{bibliography}


\end{document}